\documentclass[runningheads]{llncs}
\usepackage[T1]{fontenc}
\usepackage{graphicx}
\usepackage{marvosym}
\usepackage{makecell}
\usepackage{multirow}
\usepackage{amsmath}
\usepackage{amsfonts}
\usepackage{mathrsfs}
\usepackage[bookmarks=true]{hyperref}
\usepackage{lipsum}
\usepackage{xcolor}
\usepackage{adjustbox}
\usepackage{listings}
\usepackage{booktabs} 

\definecolor{gray}{RGB}{128,128,128}
\SetLipsumParListSurrounders{\begingroup\color{gray}}{\endgroup}

\definecolor{myred}{RGB}{255,0,0}
\definecolor{mygreen}{RGB}{0,176,80}

\definecolor{codegreen}{rgb}{0,0.6,0}
\definecolor{codegray}{rgb}{0.5,0.5,0.5}
\definecolor{codepurple}{rgb}{0.58,0,0.82}
\definecolor{backcolour}{rgb}{0.95,0.95,0.92}

\lstdefinestyle{mystyle}{
    backgroundcolor=\color{backcolour},   
    commentstyle=\color{codegreen},
    keywordstyle=\color{magenta},
    numberstyle=\tiny\color{codegray},
    stringstyle=\color{codepurple},
    basicstyle=\ttfamily\footnotesize,
    breakatwhitespace=false,         
    breaklines=true,                 
    captionpos=b,                    
    keepspaces=true,                 
    numbers=left,                    
    numbersep=5pt,                  
    showspaces=false,                
    showstringspaces=false,
    showtabs=false,                  
    tabsize=2
}

\usepackage{xspace}
\usepackage{amssymb}
\makeatletter
\DeclareRobustCommand\onedot{\futurelet\@let@token\@onedot}
\def\@onedot{\ifx\@let@token.\else.\null\fi\xspace}

\makeatother

\usepackage{color}

\begin{document}
\title{Mask-Enhanced Segment Anything Model for Tumor Lesion Semantic Segmentation
}
\titlerunning{Mask-Enhanced Segment Anything Model}
%
\author{
  Hairong Shi\inst{1, 2}  \and 
  Songhao Han\inst{1, 2}  \and 
  Shaofei Huang\inst{3}\textsuperscript{(\Letter)} \and 
  Yue Liao\inst{4} \and 
  Guanbin Li\inst{5}\textsuperscript{(\Letter)} \and 
  Xiangxing Kong\inst{6} \and  
  Hua Zhu\inst{6} \and 
  Xiaomu Wang\inst{7} \and 
  Si Liu\inst{1, 2} 
}

\authorrunning{H. Shi et al.}

\institute{Beihang University, 37 Xueyuan Road, Haidian District, Beijing, China \and Hangzhou Innovation Institute, Beihang University, 18 Chuanghui Street, Binjiang District, Hangzhou, China \and Institute of Information Engineering, Chinese Academy of Sciences, Beijing, China \\
\email{nowherespyfly@gmail.com} \and  The Chinese University of Hong Kong, Hong Kong, China \and Sun Yat-sen University, Guangzhou, China \\
\email{liguanbin@mail.sysu.edu.cn} \and Department of Nuclear Medicine, Peking University Cancer Hospital \& Institute, Peking University, Beijing, China \and Nanjing University, Nanjing, China \\}

\maketitle            
\renewcommand\thefootnote{} 
\footnote{H. Shi and S. Han -- Denote equal contribution.}
\renewcommand\thefootnote{\arabic{footnote}} 

\begin{abstract}
Tumor lesion segmentation on CT or MRI images plays a critical role in cancer diagnosis and treatment planning. 
Considering the inherent differences in tumor lesion segmentation data across various medical imaging modalities and equipment, integrating medical knowledge into the Segment Anything Model (SAM) presents promising capability due to its versatility and generalization potential.
Recent studies have attempted to enhance SAM with medical expertise by pre-training on large-scale medical segmentation datasets.
However, challenges still exist in 3D tumor lesion segmentation owing to tumor complexity and the imbalance in foreground and background regions.
Therefore, we introduce Mask-Enhanced SAM (M-SAM), an innovative architecture tailored for 3D tumor lesion segmentation. 
We propose a novel Mask-Enhanced Adapter (MEA) within M-SAM that enriches the semantic information of medical images with positional data from coarse segmentation masks, facilitating the generation of more precise segmentation masks.
Furthermore, an iterative refinement scheme is implemented in M-SAM to refine the segmentation masks progressively, leading to improved performance.
Extensive experiments on seven tumor lesion segmentation datasets indicate that our M-SAM not only achieves high segmentation accuracy but also exhibits robust generalization. The code is available at \url{https://github.com/nanase1025/M-SAM}.

\keywords{Tumor Lesion Segmentation \and Medical Image Segmentation \and Segment Anything Model}
\end{abstract}

\section{Introduction}
Tumor lesion segmentation\cite{heller2021state} aims to identify and delineate regions of abnormal tissue in medical images, \textit{e.g.}, computed tomography (CT) or magnetic resonance imaging (MRI). It plays a critical role in the processes of cancer diagnosis and treatment planning.
Since manual segmentation is extremely labor-intensive and requires a high level of expertise, deep learning based approaches~\cite{shen2017deep,litjens2017survey} have been introduced to improve efficiency and reduce the workload of physicians.
Earlier works~\cite{zhang2015deep,dou20173d} mainly rely on manually designed kernels to construct the segmentation architecture.
In recent years, the U-Net architecture~\cite{ronneberger2015u} has emerged as one of the most well-known structures for medical image segmentation~\cite{Zeng_Yang_Li_Yu_Heng_Zheng_2017,Gordienko_Gang_Hui_Zeng_Kochura_Alienin_Rokovyi_Stirenko_2019} given its effective extraction and utilization of multi-scale information.
However, it still struggles to capture long-range spatial dependencies for data with extended sequence lengths. 
Additionally, applying a U-Net model trained on one specific dataset to others with distribution shifts may lead to significant performance degradation. 
As a result, methods based on U-Net architecture often exhibit limited performances in lesion segmentation tasks, which are characterized by imbalanced foreground and background regions and require high precision.

In the field of natural image segmentation, the Segment Anything Model (SAM)~\cite{kirillov2023segment} has demonstrated outstanding versatility and impressive performances across various segmentation tasks.
It is constructed on the Transformer~\cite{vaswani2017attention} architecture, which is inherently better suited for learning long-range spatial dependencies.
Due to being trained on a vast amount of data, it also shows strong generalization capabilities in segmentation.
Furthermore, it contains an interactive system to prioritize regions of interest based on clinician cues, providing a more precise and flexible experience. 
However, due to a lack of specific knowledge in the medical imaging domain, SAM generates unsatisfying results in tasks like medical image segmentation~\cite{shi2023generalist}.

Recently, a variety of studies~\cite{ma2024segment,wang2023sam,wu2023medical,bui2023sam3d,lei2023medlsam,chen2023sam,wei2023medsam} have sought to integrate medical expertise into SAM to enhance its capabilities for medical applications.
For example, MedSAM~\cite{ma2024segment} refines the decoder with a large amount of medical data and Med-SA~\cite{wu2023medical} extends the SAM architecture using a lightweight and effective adaptation technique. 
However, these approaches involve slice-by-slice processing of volumetric images, which may lead to suboptimal performance on 3D medical images for disregarding inter-slice 3D spatial information. 
SAM-Med3D~\cite{wang2023sam}, through a comprehensive reformatting of SAM into a fully 3D architecture and extensive pre-trained on 3D medical images, has demonstrated competitive performance in general medical segmentation scenarios such as multi-organ segmentation. 
However, for the task of tumor lesion segmentation which features huge variability of tumor characteristics (\textit{e.g.}, shape, size, location, and appearance) and imbalanced tissue and tumor regions, SAM-Med3D~\cite{wang2023sam} still obtains less than satisfactory performances.

In this work, we propose a novel architecture named Mask-Enhanced SAM (M-SAM) to adapt SAM-Med3D to 3D tumor lesion segmentation tasks.
A new Mask-Enhanced Adapter (MEA) is developed to enhance the semantic information contained in image embeddings with the positional information contained in coarse segmentation masks, which further helps the generation of refined segmentation masks.
To reuse as many parameters of the pre-trained SAM-Med3D as possible, our MEA is designed to be plug-and-play and inserted between the image encoder and mask decoder for training.
Based on the MEA module, we further design an iterative refinement scheme, which leverages segmentation results from the previous iterative stage to guide the refinement process of the next iteration.
Through iterative refinement, the segmentation masks can be gradually improved, further boosting segmentation performances.
We have the following three key contributions:

(1) We introduce a novel Mask-Enhanced SAM (M-SAM) architecture to explore the application of SAM in the medical domain, validating its effectiveness in tumor lesion segmentation.

(2) We propose a Mask-Enhanced Adapter (MEA) to align the positional information of the prompt with the semantic information of the input image, optimizing precise guidance for mask prediction. Based on the design of the MEA, we further implement an iterative refining scheme to refine masks, yielding improved performances.

(3) With updates to only about $20\%$ of the parameters, our model outperforms state-of-the-art medical image segmentation methods on five tumor lesion segmentation benchmarks. Additionally, we validate the effectiveness of our method in domain transferring.


\begin{figure}[t]
   \centering
   \includegraphics[width=0.9\linewidth]{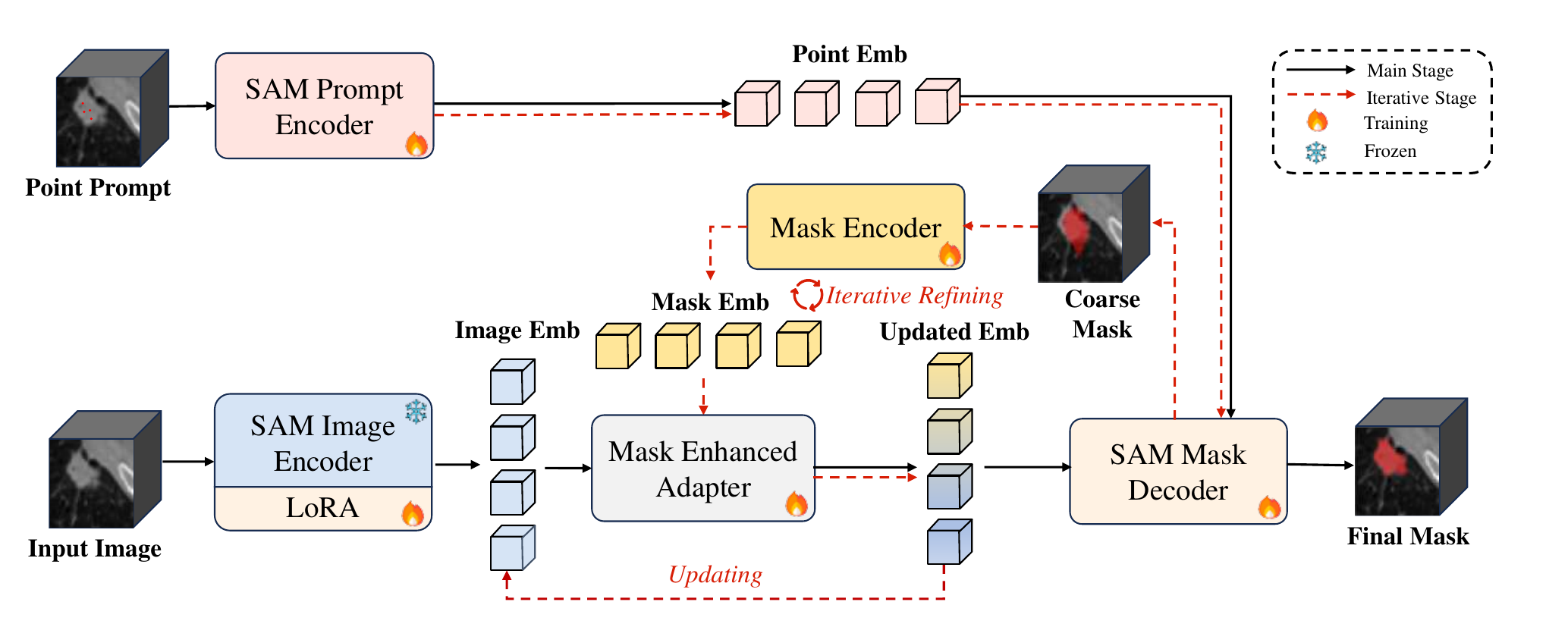}
   \caption{Overall architecture of Mask-Enhanced SAM (M-SAM). }
   \label{fig:architecture}
\end{figure}

\section{Method}
\subsection{Overall Review}
The overall architecture of our M-SAM is illustrated in Fig.~\ref{fig:architecture}. M-SAM is built upon the architecture of SAM-Med3D~\cite{wang2023sam}, which has been pre-trained on large-scale medical image segmentation datasets to obtain general medical knowledge.
It is composed of a ViT-based 3D image encoder to extract 3D image embeddings, a 3D prompt encoder to extract prompt embeddings, and a lightweight 3D mask decoder to predict segmentation results.
As shown in Fig.~\ref{fig:architecture}, the inputs are a 3D image (CT or MRI) $I \in \mathbb{R}^{C\times H\times W\times Dp}$, where $C$ represents the number of channels, $H$ the height, $W$ the width, and $Dp$ the depth (determined by the number of image modalities, \textit{e.g.}, the T1 and T2 modalities of MRI), and a randomly-initialized point as the initial point prompt.
 We first process them with SAM image encoder and SAM prompt encoder respectively, to obtain the initial image embedding $E_I^0 \in \mathbb{R}^{N_I\times D}$ and point embedding $E_P^0 \in \mathbb{R}^{N_P\times D}$. Here, $N_I$ and $N_P$ denote the dimensions of the image and point embeddings, respectively, and $D$ represents the feature dimensionality.
We also initialize a null mask $M \in \mathbb{R}^{C\times H\times W\times Dp}$ consisting of all zeros and feed it into a mask encoder to obtain the initial mask embedding $E_M^0 \in \mathbb{R}^{N_I\times D}$.
Afterward, the image embedding $E_I^0$ and mask embedding $E_M^0$ are fed into our Mask-Enhanced Adapter to obtain the updated image embedding $\hat{E}_I^0$.
By feeding $\hat{E}_I^0$ and $E_P^0$ into the SAM mask decoder, coarse segmentation mask $M^1$ can be produced, which can also be used to guide the update of image embeddings in MEA for the next stage of segmentation refinement.
The image embeddings are also updated in the iterative process accordingly. 
Through multiple stages of iterative refinement, it is possible to improve the segmentation masks in a coarse-to-fine manner continuously, thus boosting segmentation performance.

\begin{figure}[t]
   \centering
   \includegraphics[width=0.9\linewidth]{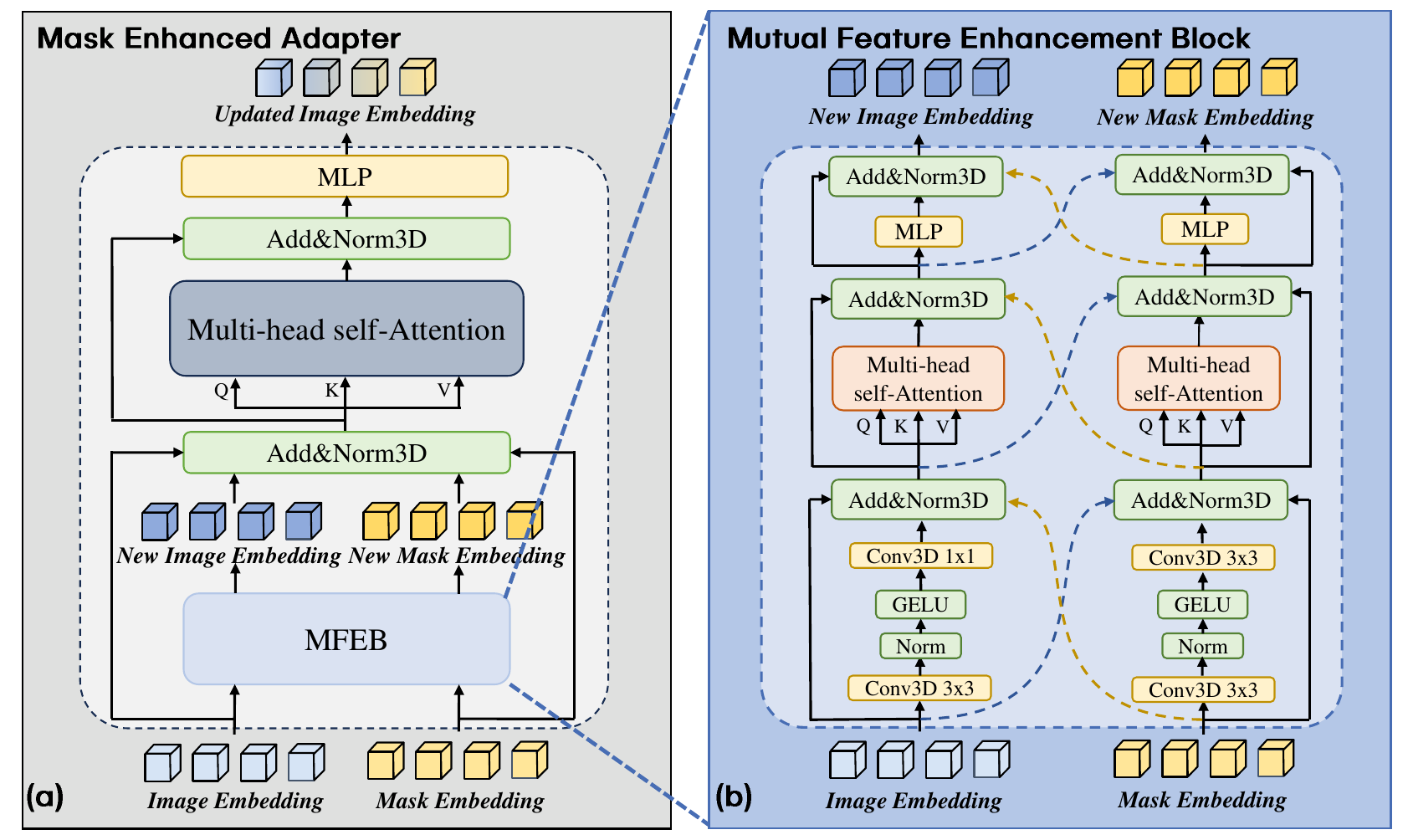}
   \caption{Module design of Mask-Enhanced Adapter. \textbf{(a)} Detailed architecture of Mask-Enhanced Adapter. \textbf{(b)} Mutual Feature Enhancement Block.}
   \label{fig:aligner}
\end{figure}

\subsection{Mask-Enhanced Adapter}
Our MEA is proposed to aggregate the image embedding with corresponding mask, so that the updated image embedding can perceive position priors of the lesion regions.
The details of the MEA are presented in Fig.~\ref{fig:aligner}(a).
For simplicity, we omit superscripts and used an iteration stage for explanation in this section.
Given the image embedding $E_I$ and the mask embedding $E_M$ as input, we first feed them into a Mutural Feature Enhancement Block (MFEB) to aggregate their features with each other.
Concretely, as shown in Fig.~\ref{fig:aligner}(b), the MFEB consists of two parallel 3D Transformer blocks, each with $E_I$ and $E_M$ as inputs.
Then, we modify the residual connections in the original Transformer block to mutual residual connections, which facilitates their fusion.
New image embedding $E'_I$ and $E'_M$ with the same shape as the original ones are produced as the output of MFEB.
Afterwards, the original image and mask embeddings are added to the new ones as residual connections, and the sum is normalized using layer normalization. 
To incorporate the mask information into the image embedding and make it aware of the foreground regions, the normalized embeddings $E''_I$ and $E''_M$ are then transmitted into a multi-head attention layer for fusion as follows:
\begin{gather}
    [Q, K, V] = (E''_I + E''_M)W_{QKV} \\
    \bar{E}_I = MHSA(Q, K, V),
\end{gather}
where $W_{QKV} \in \mathbb{R}^{D\times 3D}$ are learnable parameters of linear projections, $D$ represents the feature dimensionality and $MHSA(\cdot)$ represents Multi-Head Self-Attention module.
After adding $E''_I$ to $\bar{E}_I$ and processing them with layer normalization, we apply an MLP layer to the normalized embedding to obtain the updated image embedding $\hat{E}_I$ as the output of MEA.

\subsection{Iterative Refinement}
Based on the design of our MEA, it is possible to refine the predicted segmentation masks iteratively, thus obtaining more accurate segmentation boundaries progressively.
As shown in Fig.~\ref{fig:architecture}, the M-SAM makes predictions in a coarse-to-fine manner, beginning with the initial image embedding $E_I^0$ extracted by SAM image encoder, point embedding $E_P^0$ extracted by the SAM prompt encoder, and mask embedding $E_M^0$ derived from an all-zero mask.
We use the superscript to denote the stage number of iterative refinement in this section.
The updated image embedding $\hat{E}_I^0$ is obtained by feeding $E_I^0$ and $E_M^0$ into MEA, and then fed into SAM mask decoder with $E_P^0$ to produce the first coarse mask $M^{0}$, which is then used to generate the mask embedding $E_M^1$ for the next stage.
Concurrently, the new image embedding $E_I^1$ is updated with $\hat{E}_I^0$ and the new prompt embedding $E_{P}^{1}$ is also generated based on the last segmentation mask, yielding the refined segmentation results for the next stage.
In this way, iterative refinement is carried out continuously, resulting in increasingly accurate segmentation results.
Finally, the segmentation result of the last stage is taken as the final output of our M-SAM network.

Following SAM-Med3D~\cite{wang2023sam}, to simulate the clinical scenario of interactive segmentation, one point per iteration is randomly sampled: from the foreground in the 1st iteration, and from the error region between the coarse mask and ground truth in the subsequent 9 iteration, totaling 10 points in $10$ iterations. The updated prompt embedding $E_P$ is thus generated by feeding the newly sampled point into the prompt encoder at each iteration. This point sampling strategy is used for both training and inference in our experiments, while in real clinical use, it operates interactively with physicians.

\subsection{Loss Functions}
The overall loss function is the combination of Dice Loss~\cite{abdollahi2020vnet} and Cross-Entropy Loss with coefficients $w_0$ and $w_1$, and it is formulated as follows:
\begin{equation}
\mathcal{L}_{DiceCE} = w_0\mathcal{L}_{CE} + w_1\mathcal{L}_{Dice}, \\
\end{equation}
where
\begin{gather}
\mathcal{L}_{Dice} = 1 - \frac{2}{M} \sum_{j=1}^{M} \frac{\sum_{i=1}^{N} p_{ij} g_{ij}}{\sum_{i=1}^{N} p_{ij}^2 + \sum_{i=1}^{N} g_{ij}^2}, \\
\mathcal{L}_{CE} = -\frac{1}{N} \sum_{i=1}^{N} \sum_{j=1}^{M} g_{ij} \log(p_{ij}),
\end{gather}
where $N$ represents the number of pixels, $M$ is the number of classes, $g_{ij}$ is a binary indicator which is set to $1$ if class $j$ is the correct classification for pixel $i$, and $p_{ij}$ is the predicted probability that pixel $i$ belongs to class $j$.
$w_0$ and $w_1$ are both set as $0.5$ in our experiments.

\section{Experiments and Results}
\subsection{Experimental Setting}
\noindent\textbf{Implementation Details}.
We use the AdamW optimizer with an initial learning rate of $8e-4$ and train for $200$ epochs and the batch size was set to $4$, with a weight decay of $0.1$. In our dataset transform process, we employ a crop-or-pad strategy to standardize all images to a resolution of $128\times 128\times 128$. This involves zero-padding for images with any dimension falling short of $128$ and applying cropping for dimensions exceeding $128$. As for the data augmentation, we use RandomFlip along all three spatial axes and perform z-score normalization on each medical image data. 
All the experiments are implemented in PyTorch and trained on one NVIDIA Tesla V100 GPU.

\noindent\textbf{Datasets and Evaluation Metrics}.
We employ seven distinct segmentation tasks to comprehensively demonstrate the advantages of M-SAM. 1) Brain Tumor Segmentation Challenge 2021 (BraTS21)~\cite{baid2021rsna}, 2) Kidney Tumor Segmentation Challenge 2019 Dataset (KiTS19)~\cite{heller2021state}, 3) Medical Segmentation Decathlon Lung (MSD Lung)~\cite{antonelli2022medical}, 4) Medical Segmentation Decathlon Pancreas (MSD Pancreas)~\cite{antonelli2022medical}, 5) Liver Tumor Segmentation (LiTS)~\cite{bilic2023liver}, 6) Medical Segmentation Decathlon Hepatic (MSD Hepatic)~\cite{antonelli2022medical}, 7) Lung100*.
We adopt the Dice Similarity Coefficient (DSC) and IoU as the evaluation metrics to compare the performance of our method with other methods.
More details of the datasets and evaluation metrics can be found in our supplementary materials.


\begin{table}[t]
\caption{Comparison with U-Net-based methods and other SAM-based methods. All methods underwent parameter updates on five datasets respectively. \textbf{Bold data} indicate the highest DSC metric among all methods compared. \underline{Underline data} indicate the second highest.
}\label{tab:results}
\begin{adjustbox}{width=\textwidth}
\begin{tabular}{|r|c|c|c|c|c|c|c|c|}
\hline
\multicolumn{1}{|c|}{\multirow{2}{*}{\textbf{Method}}} & \multicolumn{1}{c|}{\multirow{2}{*}{\textbf{Cat.}}} & \multicolumn{1}{c|}{\multirow{2}{*}{\textbf{Param(M)}}} & \multicolumn{1}{c|}{\multirow{2}{*}{\makecell{\textbf{Tunable}\\ \textbf{Param(M)}}}} & \multicolumn{1}{c|}{\multirow{2}{*}{\textbf{BraTS21}}} & \multicolumn{1}{c|}{\multirow{2}{*}{\textbf{KiTS19}}} & \multicolumn{1}{c|}{\multirow{2}{*}{\textbf{Lung}}} & \multicolumn{1}{c|}{\multirow{2}{*}{\textbf{Pancreas}}} & \multicolumn{1}{c|}{\multirow{2}{*}{\textbf{LiTS}}} \\ 
 &  &  &  &  &  &  &  &  \\ \hline
TransUNet~\cite{chen2021transunet} & \multirow{5}{*}{\rotatebox{90}{\textit{U-Net-based}}} & 96 & 96 & 89.62 & 80.75 & 75.21 & 76.30 & 86.50 \\
UNETR~\cite{hatamizadeh2022unetr} & & 104 & 104 & 89.65 & 84.10 & 73.29 & 73.65 & 81.48 \\
SwinUNETR~\cite{hatamizadeh2022unetr} & & 138 & 138 & 90.48 & 87.36 & 75.55 & 70.71 & 84.00 \\
nnFormer~\cite{zhou2021nnformer} & & 151 & 151 & 90.42 & 89.09 & 77.95 & \underline{78.65} & \underline{89.83} \\
nnU-Net~\cite{isensee2018nnu} & & 16 & 16 & \underline{91.23} & \underline{89.88} & 74.31 & 76.52 & 87.97 \\ \hline \hline
Med-SA~\cite{wu2023medical} & \multirow{4}{*}{\rotatebox{90}{\textit{SAM-based}}} & 636 & 13 & 90.50 & 87.19 & 73.26 & 76.47 & 83.67 \\
SAM3D~\cite{bui2023sam3d} & & 91.88 & 1.88 & 72.90 & 80.36 & 71.42 & 71.26 & 82.27 \\
SAM-Med3D~\cite{wang2023sam} & & 101 & 101 & 86.45 & 86.65 & \underline{78.32} & 75.76 & 88.71 \\
\textbf{Ours} & & 118 & 25 & \textbf{92.08} & \textbf{93.50} & \textbf{81.62} & \textbf{80.49} & \textbf{89.95} \\ \hline
\end{tabular}
\end{adjustbox}
\end{table}

\begin{table}[t]
    \centering
    \caption{Transfer Results between datasets. \textbf{$\Delta$} represents the Dice scores difference before and after the transfer. \textbf{Rate} denotes \textbf{$\Delta$} as a proportion of the pre-transfer scores. $\textbf{*}$ represents private dataset.}
    \label{tab:transfer_comparison}
    \begin{adjustbox}{width=\textwidth}
    \begin{tabular}{|c|c|c|c|c|c|c|c|}
    \hline
    \textbf{Source} & \textbf{Target} & \textbf{Method} & \textbf{Before} & \textbf{After} & \textbf{$\Delta$} & \textbf{Rate(\%)} \\
    \hline
    \multirow{4}{*}{\textbf{LiTS Tumor}} & \multirow{4}{*}{\textbf{MSD Hepatic Tumor}} & U-Net~\cite{ronneberger2015u} & 61.54 & 50.77 & -10.77 & -17.50
 \\
    \cline{3-7}
    & & UNETR~\cite{hatamizadeh2022unetr} & 62.19 & 51.82 & -10.37 & -16.67 \\
    \cline{3-7}
    & & SAM-Med3D~\cite{wang2023sam} & 77.64 & 75.29 & \textbf{-2.35} & -3.03 \\
    \cline{3-7}
    & & \textbf{Ours} & \textbf{81.97} & \textbf{79.58} & -2.39 & \textbf{-2.92} \\
    \hline
    \multirow{4}{*}{\textbf{MSD Hepatic Tumor}} & \multirow{4}{*}{\textbf{LiTS Tumor}}  & U-Net~\cite{ronneberger2015u} & 61.08 & 52.23 & -8.85 & -14.49 \\
    \cline{3-7}
    & & UNETR~\cite{hatamizadeh2022unetr} & 64.94 & 57.15 & -7.79 & -12.00 \\
    \cline{3-7}
    & & SAM-Med3D~\cite{wang2023sam} & 85.01 & 80.82 & -4.19 & -4.93 \\
    \cline{3-7}
    & & \textbf{Ours} & \textbf{86.89} & \textbf{83.81} & \textbf{-3.08} & \textbf{-3.54} \\
    \hline\hline
    \multirow{4}{*}{\textbf{MSD Lung}} & \multirow{4}{*}{\textbf{Lung100*}} & UNETR~\cite{hatamizadeh2022unetr} & 65.16 & 52.99 & -12.17 & -18.68 \\
    \cline{3-7}
    & & U-Net~\cite{ronneberger2015u} & 73.22 & 62.32 & -10.9 & -14.89 \\
    \cline{3-7}
    & & SAM-Med3D~\cite{wang2023sam} & 78.53 & 74.35 & \textbf{-4.18} & -5.32 \\
    \cline{3-7}
    & & \textbf{Ours} & \textbf{82.17} & \textbf{77.90} & -4.27 & \textbf{-5.20} \\
    \hline
    \multirow{4}{*}{\textbf{Lung100*}} & \multirow{4}{*}{\textbf{MSD Lung}} & UNETR~\cite{hatamizadeh2022unetr} & 71.99 & 60.16 & -11.83 & -16.43 \\
    \cline{3-7}
    & & U-Net~\cite{ronneberger2015u} & 74.77 & 65.10 & -9.67 & -12.93 \\
    \cline{3-7}
    & & SAM-Med3D~\cite{wang2023sam} & 78.32 & 73.70 & -4.62 & -5.90 \\
    \cline{3-7}
    & & \textbf{Ours} & \textbf{81.62} & \textbf{77.87} & \textbf{-3.75} & \textbf{-4.59} \\
    \hline
    \end{tabular}
    \end{adjustbox}
\end{table}
\subsection{Main Result}
\noindent\textbf{Comparison with State-of-the-Art methods}.
As shown in the Table~\ref{tab:results}, our method outperforms all other methods on the aforementioned datasets. Notably, compared to the existing methods, our method achieves an average improvement of approximately $2$\%. This demonstrates the effectiveness of our method across multiple types of tumor lesion datasets. Additionally, it can be observed from the table that U-Net-based methods tend to perform better on individual datasets. However, the performance of such methods varies significantly across different datasets. For example, SwinUNETR~\cite{hatamizadeh2021swin} achieves a Dice score of over $90$\% on BraTS21, but only achieves a score of $70.7$\% on the Pancreas dataset, which is nearly $10$\% lower than our method. For other SAM-based methods, the Dice scores are generally lower than those of UNet-based methods. In terms of the number of parameters, our model has fewer parameters than the average of the methods shown in the table, and the tunable parameter count also ranks fourth among these methods. These experimental results indicate that our method can surpass these state-of-the-art methods with lower computational costs, effectively leveraging the advantages of SAM in the field of tumor lesion segmentation. Our results in Table~\ref{tab:results} are averaged across 5 runs, with Dice STD less than 0.3\% on all datasets.

\noindent\textbf{Transfer Results between Different Datasets}.
To further validate our method's generalizability, we performed transfer experiments from source to target datasets without training on the latter. We compared our method against two SAM-based and two UNet-based methods, as shown in Table~\ref{tab:transfer_comparison}. The comparison involved transfers between both public datasets and a mix of public and private datasets. Our method consistently surpassed the others, demonstrating minimal performance degradation post-transfer—3.23\% for public dataset transfers and 4.9\% for transfers involving private datasets. In contrast, UNet-based methods experienced over 10\% degradation, underscoring SAM-based methods' superior domain adaptability. Furthermore, our method outperformed SAM-Med3D~\cite{wang2023sam} in both pre- and post-transfer accuracy, with a marginal reduction in performance degradation—0.75\% and 0.72\%, respectively. This indicates our module's ability to enhance accuracy and maintain SAM's generalization capabilities across diverse datasets.

\subsection{Ablation Study}

\begin{table}[h]
    \centering
    \caption{Ablation on training setting. $\textbf{*}$ represents private dataset.}\label{tab:ablation}
    \begin{adjustbox}{width=\textwidth}
    \begin{tabular}{|c|c||cc|cc||cc|}
        \hline
        \multirow{2}{*}{\textbf{Setting Variants}} & \multirow{2}{*}{\makecell{\textbf{Tunable}\\ \textbf{Param(\%)}}} & \multicolumn{2}{c|}{\textbf{KiTS19}} & \multicolumn{2}{c||}{\textbf{BraTS2021}} & \multicolumn{2}{c|}{\textbf{Lung100*}} \\ \cline{3-8} 
         & & \multicolumn{1}{c|}{\textbf{DSC}} & \textbf{IoU} & \multicolumn{1}{c|}{\textbf{DSC}} & \textbf{IoU} & \multicolumn{1}{c|}{\textbf{DSC}} & \textbf{IoU} \\ \hline
        baseline & 0 & \multicolumn{1}{c|}{73.28} & 57.82 & \multicolumn{1}{c|}{83.30} & 71.38 & \multicolumn{1}{c|}{10.07} & 5.30 \\
        \hline\hline
        w/o MEA & 8.09 & \multicolumn{1}{c|}{86.75} & 76.59 & \multicolumn{1}{c|}{84.35} & 72.94 & \multicolumn{1}{c|}{77.96} & 63.88 \\\hline
        backbone full-ft & 100 & \multicolumn{1}{c|}{\textbf{92.67}} & \textbf{86.34} & \multicolumn{1}{c|}{\textbf{92.15}} & \textbf{85.44} & \multicolumn{1}{c|}{82.09} & 69.62 \\ \hline
        \textbf{M-SAM(Ours)} & 21.41 & \multicolumn{1}{c|}{92.58} & 86.19 & \multicolumn{1}{c|}{92.08} & 85.32 & \multicolumn{1}{c|}{\textbf{82.17}} & \textbf{69.74} \\ \hline
    \end{tabular}
    \end{adjustbox}
\end{table}

We conduct ablation study on KiTS19, BraTS2021 and Lung100* datasets to validate the effectiveness of our method.
As illustrated in Table~\ref{tab:ablation}, after fine-tuning with LoRA, our method requires updating only 21\% of the parameters to achieve results nearly identical or even better than full-parameter fine-tuning. This efficient fine-tuning approach significantly reduces the requirements for computational resources. 
With the incorporation of our proposed Mask-Enhanced Adapter module, Dice and IoU metrics on public datasets further increase by over $3.5$\% and $6$\%, respectively. On our private Lung100* dataset, the performance improvement reaches approximately $2.8$\% for Dice Score and $4$\% for IoU respectively.

\section{Conclusion}
In this work, we introduce Mask-Enhanced SAM (M-SAM), a novel architecture designed specifically for 3D tumor lesion segmentation. In M-SAM, the Mask-Enhanced Adapter (MEA) enhances the semantic information of medical images by incorporating positional data from coarse segmentation masks, which assists in generating more precise segmentation masks. Additionally, we implement an iterative refinement scheme in M-SAM to progressively improve segmentation mask quality, resulting in enhanced performance. Extensive experiments on seven tumor lesion segmentation datasets demonstrate that M-SAM outperforms existing state-of-the-art methods and robust generalization capabilities of the proposed architecture.\\

\noindent\textbf{Acknowledgments.}
This research is supported in part by National Natural Science Foundation of China (NO. 62122010, U23B2010 and NO.~62322608), Zhejiang Provincial Natural Science Foundation of China (Grant NO.~LDT23F02022F02), Key Research and Development Program of Zhejiang Province (Grant 2022C01082) and Beijing Natural Science Foundation (NO.~L231011).

\bibliographystyle{splncs04}
\bibliography{reference}

\end{document}